\documentclass{ws-procs9x6}
\usepackage{dpscolor}
\begin{document}
\newcommand{\rubin}{\textColor{RubineRed}}
\newcommand{\black}{\textColor{Black}}

\title{The ANTARES Experiment:\\Past, Present and Future}

\author{Igor Sokalski\\
on behalf of the ANTARES Collaboration~\footnote{http://www.slac.stanford.edu/spires/find/experiments/www2?expt=ANTARES}}

\address{INFN / Bari\\via Amendola 173\\I-70126 Bari\\Italia\\
E-mail: Igor.Sokalski@ba.infn.it}

\maketitle

\abstracts{The {\sc Antares} collaboration aims to build a deep underwater 
Cherenkov neutrino telescope in the Mediterranean Sea at 2500 m depth, about 
40 km off-shore of La Seyne sur Mer, near Toulon (42$^{\circ}$50$^{'}$\,N, 
6$^{\circ}$10$^{'}$\,E). The collaboration was formed in 1996 and the 
experiment is currently in the construction phase. The final {\sc Antares} 
detector, consisting of 12 strings each equipped with 75 photomultiplier tubes,
is planned to be fully deployed and taking data by 2007. The project aims to 
detect atmospheric and extraterrestrial neutrinos with energies above 
$E_{\nu} \sim$\,10\,GeV by means of the Cherenkov light that is generated in 
water by charged particles which are produced in the neutrino interactions.}

\rubin

\vspace{-136.0mm}

\begin{center}
\begin{tabular}{cc}
\rule{50mm}{0mm} & {\footnotesize {\sc Invited talk given}}\\ 
\rule{50mm}{0mm} &
{\footnotesize {\sc at the 44th Workshop on `QCD at Cosmic Energies:}}\\
\rule{50mm}{0mm} &
{\footnotesize {\sc the Highest Energy Cosmic Rays and QCD',}}\\
\rule{50mm}{0mm} & 
{\footnotesize {\sc Erice, Italy, Aug 29 - Sep 5, 2004}}\\
\rule{50mm}{0mm} & 
{\footnotesize {\sc (to be published in the Proceedings)}}\\
\end{tabular}
\end{center}

\vspace{-68.6mm}

\begin{picture}(510,196)
\put(252,40){\oval(224,69)}
\end{picture}

\vspace{99mm}

\black

\section{Introduction}
Neutrinos are an attractive tool for astrophysical investigations since they 
are weakly interacting and hence they are not absorbed in sources or during 
propagation to the  Earth (moreover, being neutral, they are not deflected by 
magnetic fields). Nevertheless, due to the same property, huge volume neutrino 
detectors are needed. {\sc Antares} is one of the several on-going
projects~{\Large$^{1-6}$} on underwater/ice neutrino telescopes.

The idea to construct large deep underwater Cherenkov detectors for neutrino
astronomy using natural basins was formulated by M.~A.~Markov in 
1960~\cite{markov} as an alternative to the underground neutrino telescopes
whose sensitivity does not allow to detect neutrinos from the cosmic
accelerators. The effective area of the cosmic neutrino detector must be at 
least of the order of 0.1-1\,km$^{2}$ which goes far beyond the constructional 
possibilities of the underground technique. The underwater/ice neutrino 
telescopes aim to detect neutrinos with energies above $E_{\nu} \sim$\,10\,GeV 
by means of the detection of the Cherenkov light that is generated in water by 
charged particles which are produced in neutrino interactions. Water (or ice) 
serves for three purposes: it represents {\it i)} a shield which protects from
atmospheric muon background, {\it ii)} a target in which neutrino interaction
occurs and {\it iii)} the detection medium where the Cherenkov light is emitted
and propagates. The Cherenkov photons are detected by a 3D grid of 
photomultipliers (PMTs) immersed deep (at the km scale depth) under water or
ice. The typical detector consists of vertical lines (strings) equipped with
PMTs and spaced by 10-100\,m from each other. The neutrino direction can be
reconstructed using times and position of hit PMTs, the energy is
estimated using the amplitude information. 

As a matter of fact, neutrinos oscillate during their propagation from sources,
hence signal from sources should consist of all neutrino flavors. However, 
neutrino telescopes were originally optimized for the detection of muon 
neutrinos through the reaction $\nu_{\mu}\,N\stackrel{CC}{\rightarrow}\mu\,X$, 
since muons propagate through large distances increasing the 'effective' target
mass of the detector (for 1\,TeV muon, {\it e.g.}, the mean range is about 
2.5\,km\,w.e.). Moreover, muon tracks can be reconstructed with good angular 
resolution allowing to point-back to neutrino sources. On the other hand, 
detection of shower events (which are produced in the detector sensitive volume
by $\nu_{e}$'s, $\nu_{\tau}$ with energies lower than 
$E_{\nu_{\tau}}\sim1$\,PeV and neutral current interactions) has also 
interesting potentials, as discussed, {\it e.g.}, in \cite{amsh,reno1}, as 
well as detection of extremely high energy $\nu_{\tau}$'s through 
$\tau$-leptons  whose ranges at PeV and EeV energy range are compatible with 
muon ones (see, {\it e.g.}, \cite{bang,bmss}). In a narrow energy range around 
6.33\,PeV the resonant processes $\bar\nu_{e}\,e^{-}\to W^{-} (Z^{\circ})\to X$
(where $X$ can be $\bar\nu_{e}\,e^{-}$, $\bar\nu_{\mu}\,\mu^{-}$, 
$\bar\nu_{\tau}\,\tau^{-}$ or hadrons)~{\Large$^{12-14}$} which result both 
in `shower' and `track' events must be taken into account, as well.

The {\sc Antares}~\footnote{{\sc Antares} $=$ \underline{{\sc A}}stronomy with 
a \underline{{\sc N}}eutrino \underline{{\sc T}}elescope and
\underline{{\sc A}}byss environmental \underline{{\sc Res}}earch.} 
Collaboration  was formed in 1996. By that time the pioneering underwater
neutrino project {\sc Dumand} in the Pacific Ocean~\cite{dumand} had been 
cancelled, the {\sc Baikal} detector in the Siberian lake Baikal had reported 
the first atmospheric neutrinos detected underwater~\cite{baikal,baikal_1}, 
the {\sc Amanda-B}~\cite{amandab} detector at the South Pole was under 
construction ({\sc Amanda-A}~\cite{amandaa} had been taking data for 2 years) 
and the {\sc Nestor} experiment~\cite{nestor} in the Mediterranean See near 
Pilos (Greece) was at the R\&D phase. 

Now the {\sc Antares} project joins about 200 scientists and engineers from 
France, Germany, Italy, Russia, Spain, The Netherlands and the United Kingdom. 
After an extensive R\&D program the collaboration entered the phase of 
construction of a 12-string neutrino telescope in the Mediterranean Sea at 
2500\,m depth. Given the presence of {\sc Amanda}~\cite{amanda} and 
{\sc IceCube}~\cite{icecube} at the South Pole, a detector in the 
Mediterranean will allow to cover the whole sky looking for extraterrestrial 
neutrino sources.

\section{The Past}
\subsection{Site investigation}
In 1996-99 an intense R\&D program was performed. The deployment and recovery
technologies, electronics and mechanical structures were developed and tested
with more than 30 deployments of autonomous strings. The environmental
properties at the detector site were investigated~{\Large$^{19-21}$}.

Concerning the optical backgrounds it was found that  PMT counting rates have
very strong temporal variations due to bioluminiscence organisms (see
Figure~1). The baseline rate slowly varies between 50 kHz and 300 kHz on a
10$^{``}$ PMT being accompanied by short (from several seconds to several
minutes) several hundred kHz bioluminiscence `bursts'. Nevertheless, the
measured optical background is 50-70\% of the time below 100 kHz which is a
rate acceptable for data taking. To suppress the high optical backgrounds, 
coincidence requirements and PMT thresholds higher than 0.25 photo-electrons 
will be used during data taking, which will slightly reduce the efficiency at 
low energies without deteriorating it in the region of interest for 
astrophysical sources ($E_{\nu}>$ 100 GeV).

Light transmission loss for glass containers that house PMTs was found strong
in long-term tests for up-looking surfaces. It led to the decision to turn all
PMTs downward. Signal loss due to bio-fouling and sedimentation was measured
to be 1.6\% after 8 months at equator of glass sphere saturating with time.

The optical properties of water at the experiment site were measured during
several years. The effective attenuation length varies in a range
40\,m\,\,$<\Lambda_{att}<$\,\,60\,m (see Figure~2) while the effective 
scattering length is $\Lambda_{scatt}>$\,\,200\,m for blue light 
($\lambda=$\,466 nm). Only 5\% of the photons emitted by an isotropic source 
located 24\,m from a PMT are collected out of a 10\,ns time window being 
delayed due to scattering. This allows a good time resolution needed for event 
reconstruction.

\begin{figure}[h]
\centerline{\epsfxsize=7.5cm\epsfbox{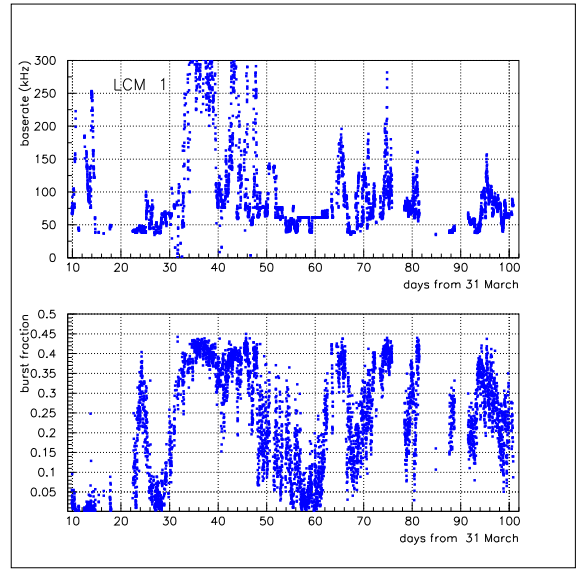}}
\caption{
Summary of counting rate in 3 PMTs during 65 days in April--May, 2003 vs time.
Top panel: the average baseline rate. Bottom panel: the fraction of time the
rate is significantly higher than this average baseline rate (burst fraction).
}
\end{figure}
\begin{figure}[h]
\centerline{\epsfxsize=6.7cm\epsfbox{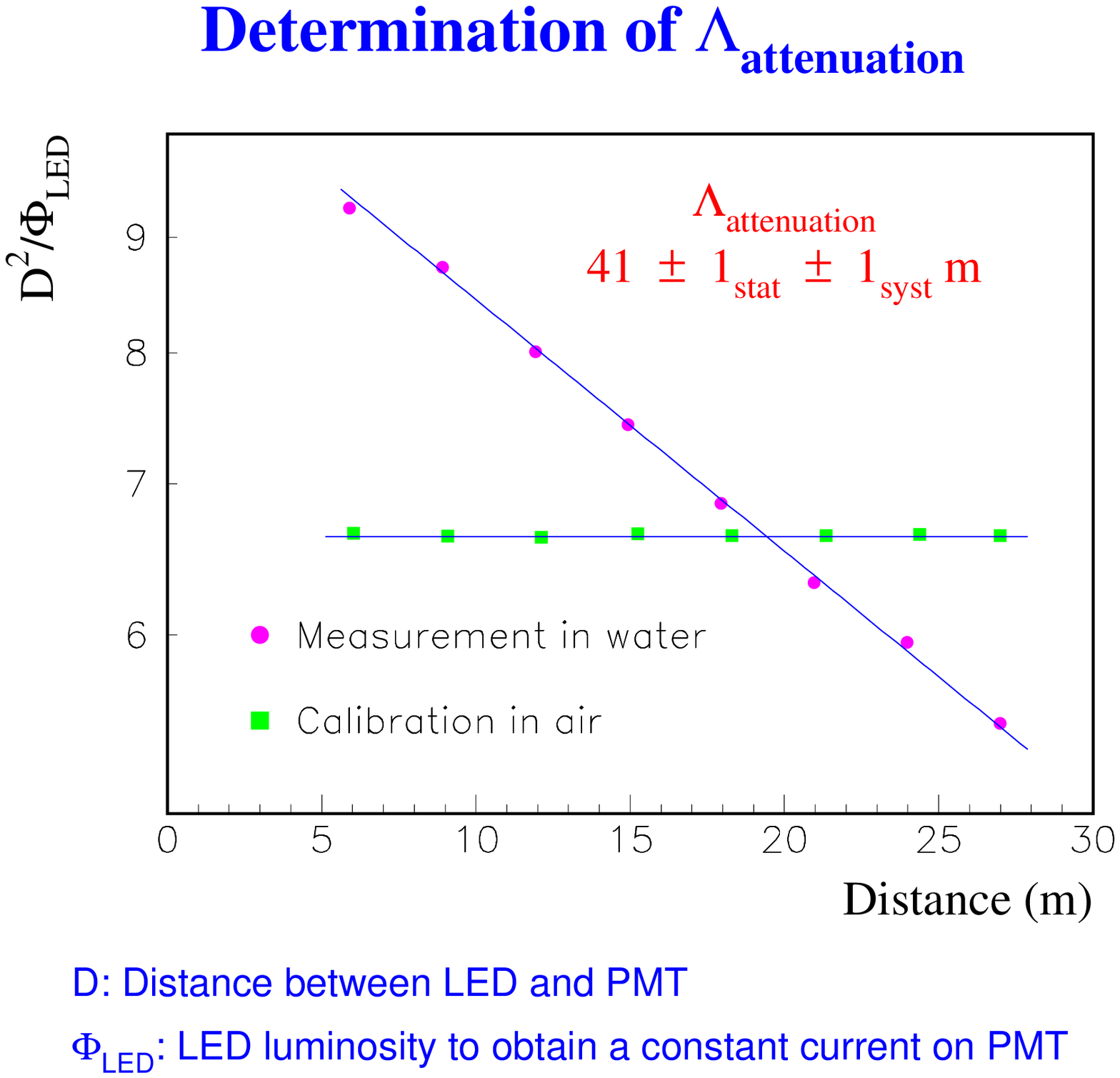}}
\caption{
Determination of the effective attenuation length from the setup immersed in 
December 1997.
}
\end{figure}

\subsection{Prototype strings}
Several prototype strings were immersed deep in the sea and operated in the
frame of the {\sc Antares} R\&D program.

Firstly, a 350\,m length `demonstrator string'  instrumented with 7 PMTs was
deployed at a depth of 1100\,m, 40\,km off the coast of Marseille and operated
for 8 months (November 1999 - July 2000). The string was controlled and read
out via a $\sim$40\,km-long electro-optical cable connected to the shore 
station. It
allowed to test the deployment procedure with a full-scale string, positioning
system and collect $\sim$5$\cdot$10$^{4}$ seven-fold coincidences from
atmospheric muons. Relative distances were measured with an accuracy of
$\sim$5\,cm and accuracy of absolute positioning was $\sim$1\,m. The angular
distribution of atmospheric muons was reproduced reasonably (Figure~3) and the
fraction of multi-muon events was found to be about 50\% which is in agreement
with expectation for such a shallow depth as 1100\,m.
\begin{figure}[h]
\centerline{\epsfxsize=7.5cm\epsfbox{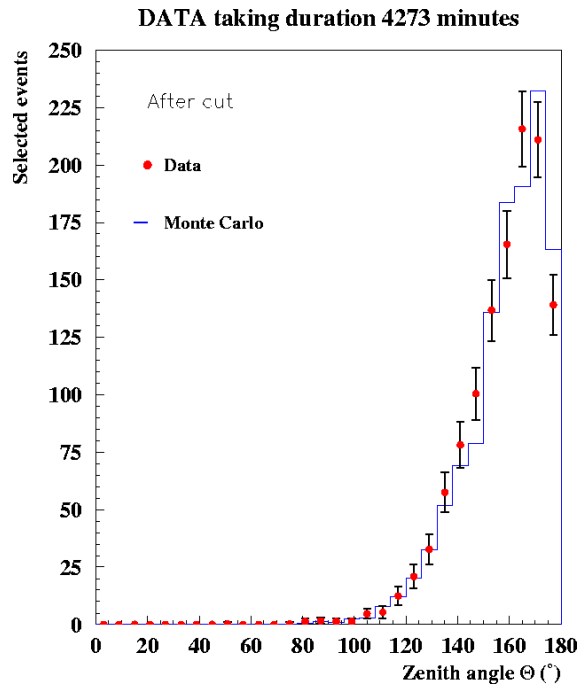}}
\caption{
Reconstructed angular distribution from the `demonstrator string' data. The 
areas of the curves are normalized to each other.
}
\end{figure}

Then in December 2002 and February 2003 the prototype and the mini
instrumentation lines~\cite{prot} have been deployed and positioned on sea bed
within a few meters from their nominal positions. In March 2003 a manned
submarine `Nautile' successfully connected both lines to the electro-optical 
cable. The data have been taken continuously until the recovery of the 
prototype line in July 2003 and analyzed to study the optical background at 
the {\sc Antares} site. Two problems occurred in the prototype tests. A water 
leak developed in one of the electronic containers due to a faulty supplier 
specification for  a connector. This made further operation impossible and the 
line was recovered in May 2003. Also, a defect in the clock signal transmission
caused by a broken optical fiber inside the line meant that data with timing 
information at nanosecond precision were unavailable. Nevertheless, during 
about 100 days of the prototype line operation data were recorded, both on the 
functionality of the detector and on environmental conditions.

\section{The Present}
\subsection{The ANTARES 12-string detector: design}
After R\&D experience, the collaboration moved to the next stage: construction
of a 12-string detector which can be considered as a step toward a 1\,km$^{3}$
detector. The design of the planned detector is shown in Figure~4.
\begin{figure}[b]
\centerline{\epsfxsize=11.4cm\epsfbox{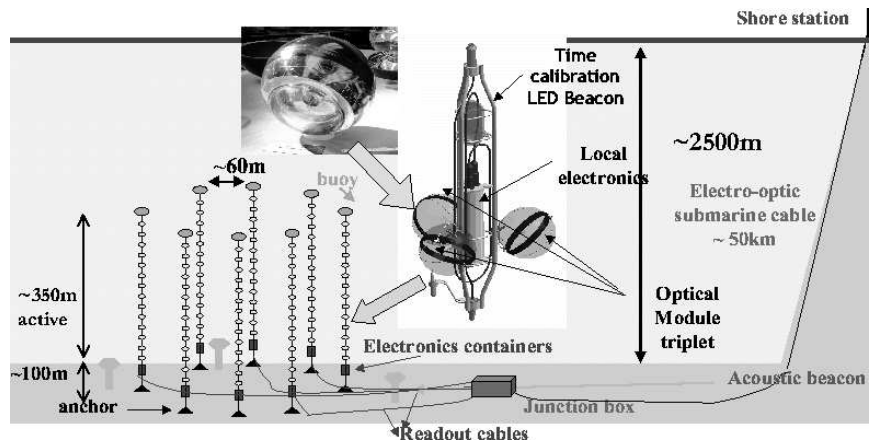}}
\caption{
Schematic view of the {\sc Antares} 12-string detector.
}
\end{figure}
Strings are anchored at the sea floor and held taut by buoys. Each string is
instrumented with 75 optical modules (OMs)~\cite{OM} containing 10$^{``}$
Hamamatsu R7081-20 PMTs~\cite{PMT} housed in glass spheres. The main 
specifications that {\sc Antares} PMTs must satisfy are: a transit time spread 
of less than 3 ns (FWHM), a peak to valley ratio larger than 2, a dark count of
less than 20 kHz for a 0.25 photo-electron threshold, and a gain larger than 
$5 \cdot 10^7$ for high voltage lower than 2000 V. Laser calibrations in a dark
room using a prototype line made of 5 storeys have shown that, after 
corrections of clock delays between consecutive storeys, the achievable timing 
resolution is $\sigma \sim 0.9-1.2$ ns. OMs are grouped in triplets at 25 
levels separated by 14.5\,m. 3 PMTs in each triplet are oriented at 
45$^{\circ}$ to the nadir. Strings are separated from each other by 
$\sim$60--70\,m. All the strings are connected to a `junction box' by means of 
electro-optical link cables. The junction box is connected to the shore station
by a 50\,km long 48-fiber electro-optical cable (which was deployed in October 
2001). PMT signals are processed by Analogue Ring Samplers which measure the 
arrival time and charge for 1\,p.e.-pulses (99\% of the pulses) and perform 
wave form digitization for larger amplitudes. Digitized data from each OM are 
sent to shore ($\sim$1\,GB/s/detector). The data flow is reduced down to 
$\sim$1\,MB/s on the shore by means of an on-shore data filter~\cite{data}. A 
100 PC farm is foreseen on shore to process and collect the data. The telescope
will be complemented with an instrumentation string for hydrological parameter 
measurements and for calibration purposes. The deployment of the detector is 
planned for 2005-2007.

\subsection{The 12-string detector: physical performance}
The main background for deep underwater Cherenkov experiments comes from
atmospheric muons. To suppress this background one selects the `$\nu_{\mu}$ 
events' only out of events that have been reconstructed as up-going ones.

Figure~5 shows the neutrino effective area for the 12-string {\sc Antares} 
detector for different nadir angles. The neutrino effective area is the 
sensitive area of the detector `seen' by neutrinos producing detectable muons 
when entering the Earth. Since cross section of $\nu$-$N$ charged current 
interactions (in which muons are produced) is very small, the neutrino 
effective area is orders of magnitude lower compared to geometrical dimensions 
of the detector. The decrease of the area in the vertical region at high 
energies is due to the Earth shadowing effect which is remarkable for neutrino 
energies larger than $E_{\nu}\sim$\,50\,TeV (see, {\it e.g.}, \cite{lms}).

\begin{figure}[h]
\centerline{\epsfxsize=9.1cm\epsfbox{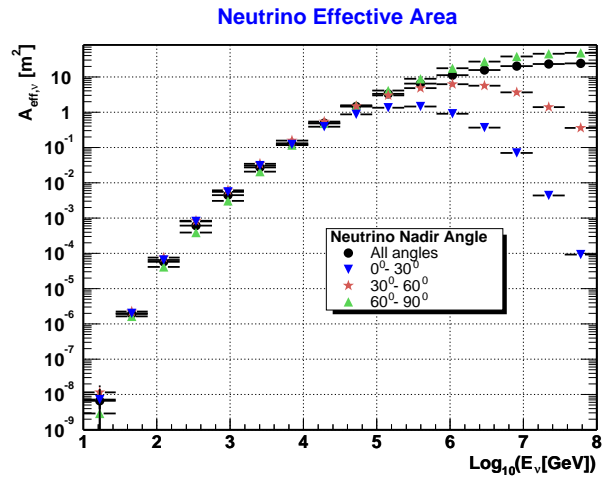}}
\caption{
The neutrino effective area for 12-string {\sc Antares} detector computed for
different nadir angles vs neutrino energy.
}
\end{figure}
\begin{figure}[h]
\centerline{\epsfxsize=8.8cm\epsfbox{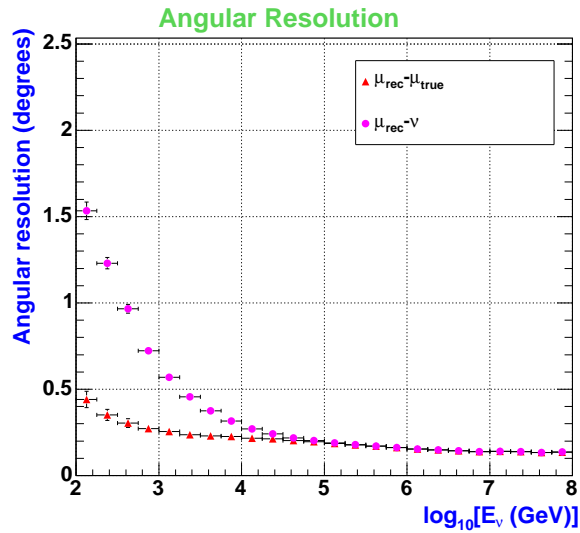}}
\caption{
{\sc Antares} angular resolution vs neutrino energy. The dots: median angle
between the simulated neutrino and reconstructed muon direction. Triangles:
the median of the angle between the simulated muon and reconstructed one.
}
\end{figure}

\begin{figure}[h]
\centerline{\epsfxsize=9.5cm\epsfbox{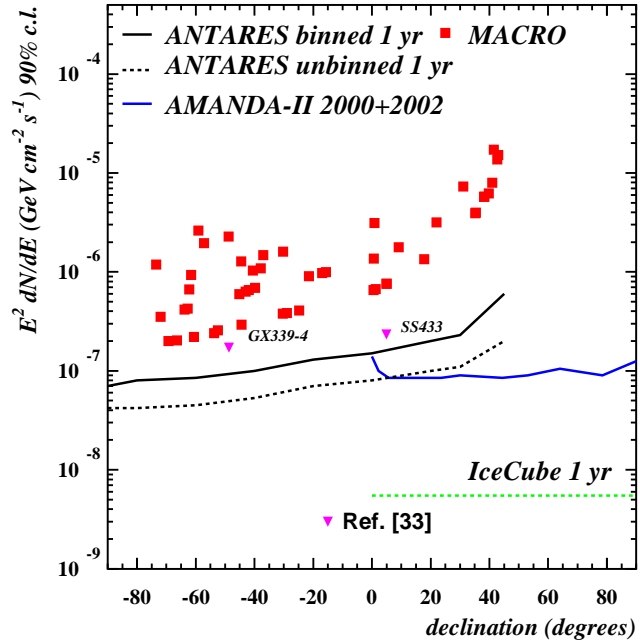}}
\caption{
Upper limits (90\% CL) on $E^{-2}$ neutrino fluxes as a function of the source
declination for {\sc Macro}~\protect\cite{MACRO_pls}, expected sensitivity of
{\sc Amanda-II} corresponding to 2000-2002 data~\protect\cite{AMANDA_pls},
{\sc IceCube}~\protect\cite{IceCube_pls} and
{\sc Antares}~\protect\cite{ANTARES_pls} (for a search method using a grid in
the sky and an unbinned method based on likehood ratio). The triangles
indicates the expected neutrino flux from two persistent micro-quasars as
calculated in \protect\cite{distefano}.
}
\end{figure}

The angular resolution as obtained with the {\sc Antares} simulation is shown 
in Figure~6. The plot shows the median angle between the neutrino source and 
the reconstructed muon (that is the pointing capability for a neutrino source) 
and the `intrinsic angular resolution', that is the angle between the `true' 
muon track and reconstructed one. The angular resolution of the {\sc Antares} 
detector is about 0.2$^{\circ}$ for $E_{\nu} \ge$\,10\,TeV where it is limited 
only by PMT transit time spread and light scattering and 
$\sim$0.3$^{\circ}$--1.5$^{\circ}$ at $E_{\nu} \sim$\,0.1--10\,TeV where
accuracy is dominated by $\nu - \mu$ kinematics.

\begin{figure}[h]
\centerline{\epsfxsize=9.6cm\epsfbox{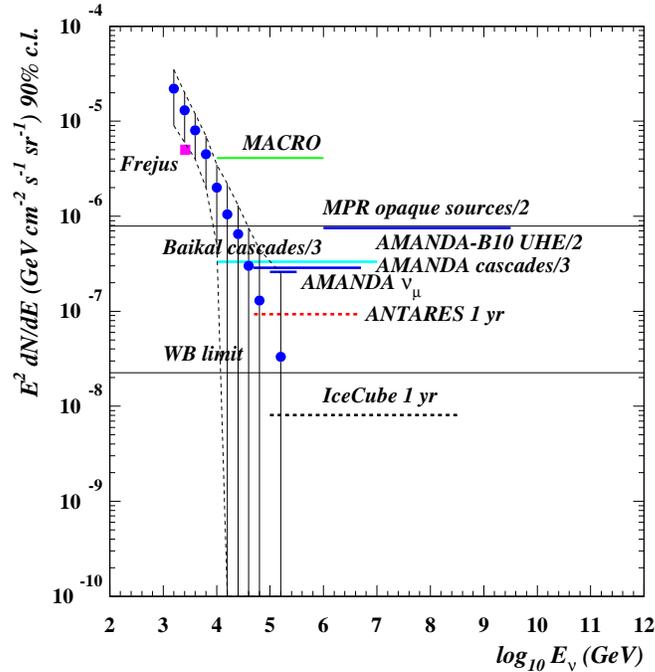}}
\caption{
Sensitivity for {\sc Antares} to the diffuse neutrino flux compared to 90\% 
c.l. limits on diffuse $E^{-2}$ fluxes of $\nu_{\mu}+\bar{\nu}_{\mu}$ in the 
hypothesis of $\nu$ oscillations as measured by
{\sc Amanda-II}~\protect\cite{AMANDA_dif},
{\sc Baikal}~\protect\cite{Baikal_nu}, {\sc Macro}~\protect\cite{MACRO_nu}
and {\sc Frejus}~\protect\cite{frejus}.
Limits for other flavors than $\nu_{\mu}$ (cascades) have been divided by the
number of contributing flavors. Sensitivity of
{\sc IceCube}~\protect\cite{IceCube_pls} is shown, as well. Dots are the
measured atmospheric neutrino flux by {\sc Amanda-II}~\protect\cite{AMANDA_nu}.
Waxmann \& Bahcall limit on diffuse neutrino flux~\protect\cite{wb} and
Mannheim-Protheroe-Rachen limit on diffuse neutrino flux for opaque
sources~\protect\cite{mpr} are given by thin horizontal lines.
}
\end{figure}

The sky coverage for the {\sc Antares} detector is 3.5$\pi$ sr with the whole 
Southern hemisphere observable. Promising neutrino source candidates as, 
{\it e.g.}, the Galactic Center and supernova remnant
RX~J1713.7-3946~\cite{cang} are visible 67\% and 78\% of the time,
correspondingly. The instantaneous overlap with South Pole detectors
{\sc Amanda}~\cite{amanda} and {\sc IceCube}~\cite{icecube} is 0.5$\pi$ sr with
1.5$\pi$ sr common view per day.

The sensitivity of {\sc Antares} for 1 year of data taking as a function of
declination for two different methods of searches for point-like sources is
shown in Figure~7, where it is compared to other experiments. It can be seen
that there is a real hope to detect a signal from the most promising sources
({\it e.g.}, galactic micro-quasars~\cite{distefano}).

Studies on the sensitivity of {\sc Antares} to a diffuse muon neutrino flux
from populations of sources have brought to the result given in Figure~8. The
sensitivity of the detector to diffuse neutrino fluxes allows to reach
Waxmann \& Bahcall limit~\cite{wb} in about 4 years.

\begin{figure}[h]
\centerline{\epsfxsize=9.1cm\epsfbox{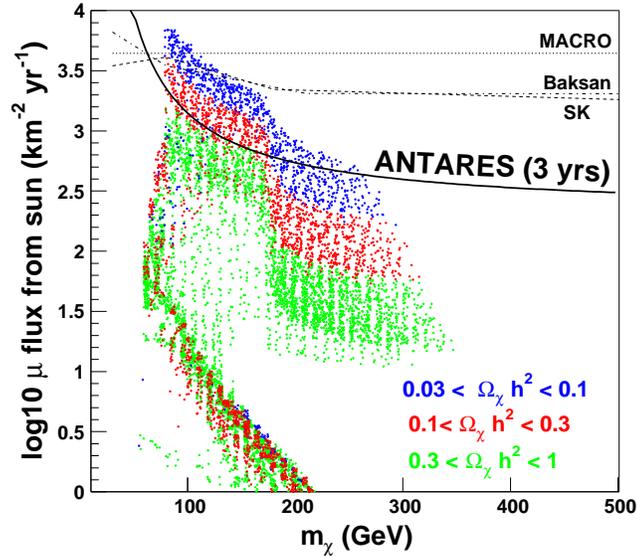}}
\caption{
The {\sc Antares} 12-string detector sensitivity to the muon flux from
neutralino annihilations in the core of the Sun in comparison to the upper
limits from {\sc MACRO}~\protect\cite{MACRO_neu},
{\sc Baksan}~\protect\cite{Baksan_neu},
{\sc Super-Kamiokande}~\protect\cite{SK_neu} and to predictions of
mSUGRA~\protect\cite{sugra} models.
}
\end{figure}

Apart from the neutrino astronomy, other motivations are included in the
{\sc Antares} scientific program. Neutralinos, the best candidates for cold
dark matter, can be gravitationally captured in the massive astrophysical
objects such as the Sun, the Earth or the Galactic Center and annihilate there
producing neutrinos in the decay chain. Expected {\sc Antares} sensitivity to
the muon flux from neutralino annihilation in the core of the Sun for the case
of a `hard' neutrino spectrum (assuming 100\% annihilations
$\chi\,\bar\chi\,\to\,W^{+}\,W^{-}$) is shown in Figure~9.

\section{The Future}
The {\sc Antares} 12-string detector can be considered as a first stage toward
a km$^3$-scale telescope, in which European institutions involved in current
deep underwater neutrino telescope projects ({\sc Antares}~\cite{antares},
{\sc Nemo}~\cite{nemo} and {\sc Nestor}~\cite{nestor}) are already
collaborating. This network, {\sc KM3NeT}~\cite{km3}, will give birth to a
1\,km$^{3}$ Northern hemisphere telescope which will complement the 
{\sc IceCube} detector~\cite{icecube} in the Southern hemisphere.

\section{Conclusions}
The construction of the {\sc Antares} detector is well underway. It is planned
to be fully deployed and start to take data in 2007. Calculations based on the
data on environmental conditions at the experiment site and on studied
properties of electronic components shows that predicted sensitivity of the
detector to diffuse neutrino fluxes, point-like neutrino searches and WIMP
searches is quite enough to participate in a competition with other
experimental groups. The deployment of the {\sc Antares} neutrino telescope
can be considered as a step toward the deployment of a 1 km$^3$ detector in
the Mediterranean Sea.


\begin{thebibliography}{99}
\bibitem{amanda}
The {\sc Amanda} experiment: E.~Andres {\it et al.}, {\it Astropart. Phys.}
{\bf 13}, 1 (2000) (available from $<$arXiv:{\it astro-ph/9906203}$>$); 
J.~Ahrens {\it et al.}, e-print {\it astro-ph/0409423}. See also URL 
{\it http://amanda.uci.edu/}
%%CITATION = ASTRO-PH 9906203;%%
%%CITATION = ASTRO-PH 0409423;%%
\bibitem{antares}
The {\sc Antares} experiment: E.~Aslanides {\it et al.}, e-print
{\it astro-ph/9907432}; E.~V.~Korolkova {\it et al.}, 
{\it Nucl. Phys. Proc. Suppl.} {\bf 136}, 69 (2004) (available from 
$<$arXiv:{\it astro-ph/0408239}$>$). See also URL 
{\it http://antares.in2p3.fr/}
%%CITATION = ASTRO-PH 9907432;%%
%%CITATION = ASTRO-PH 0408239;%%
\bibitem{baikal}
The {\sc Baikal} experiment: I.~A.~Belolaptikov {\it et al.},
{\it Astropart. Phys.} {\bf 7}, 263 (1997); Ch.~Spiering {\it et al.},
{\it Nucl. Phys. Proc. Suppl.} {\bf 138}, 175 (2005). See also URL 
{\it http://baikal1.jinr.ru/}
%%CITATION = APHYE,7,263;%%
\bibitem{icecube}
The {\sc IceCube} experiment: J.~Alvarez-Muniz and F.~Halzen,
{\it AIP Conf. Proc.} {\bf 579}, 305 (2001) (available from 
$<$arXiv:{\it astro-ph/0102106}$>$); Sh.~Yoshida {\it et al.},
{\it Nucl. Phys. Proc. Suppl.} {\bf 138}, 179 (2005). See also URL 
{\it http://icecube.wisc.edu/}
%%CITATION = ASTRO-PH 0102106;%%
\bibitem{nemo}
The {\sc Nemo} experiment: C.~N.~De~Marzo {\it et al.}, in Proceedings of the
Vulcano Workshop 2000: Frontier Objects in Astrophysics and Particle Physics,
Vulcano, Italy, 22-27 May 2000, p.\,593; E.~Migneco {\it et al.},
{\it Nucl. Phys. Proc. Suppl.} {\bf 136}, 61 (2004). See also URL
{\it http://nemoweb.lns.infn.it/}
%%CITATION = NUPHZ,136,61;%%
\bibitem{nestor}
The {\sc Nestor} experiment: L.~K.~Resvanis {\it et al.},
{\it Nucl. Phys. Proc. Suppl.} {\bf 35}, 294 (1994); L.~K.~Resvanis 
{\it et al.}, {\it Nucl. Phys. Proc. Suppl.} {\bf 138}, 187 (2005). See also 
URL {\it http://www.nestor.org.gr/}
%%CITATION = NUPHZ,35,294;%%
\bibitem{markov}
M.~A.~Markov, in Proceedings of 10th International Conference on High Energy
and Nuclear Physics, Rochester, USA, 25 August - 1 September 1960, p.\,579.
\bibitem{amsh}
M.~Ackermann {\it et al.} (The {\sc Amanda} Collaboration), 
{\it Astropart. Phys.} {\bf 22}, 127 (2004) (available from 
$<$arXiv:{\it astro-ph/0405218}$>$).
%%CITATION = ASTRO-PH 0405218;%%
\bibitem{reno1}
S.~I.~Dutta, M.~H.~Reno and I.~Sarcevic , {\it Phys.\,Rev.} {\bf D62}, 123001 
(2000) (available from $<$arXiv:{\it hep-ph/0005310}$>$).
%%CITATION = HEP-PH 0005310;%%
\bibitem{bang}
J.~G.~Learned and S.~Pakvasa, {\it Astropart.\,Phys.} {\bf 3}, 267 (1995)
(available from $<$arXiv:{\it hep-ph/9405296}$>$).
%%CITATION = HEP-PH 9405296;%%
\bibitem{bmss}
E.~Bugaev, T.~Montaruli, Yu.~Shlepin and I.~Sokalski, {\it Astropart. Phys.} 
{\bf 21}, 491 (2004) (available from $<$arXiv:{\it hep-ph/0312295}$>$).
%%CITATION = HEP-PH 0312295;%%
\bibitem{glashow1}
S.~L.~Glashow, {\it Phys.\,Rev.} {\bf 118}, 316 (1960).
%%CITATION = PHRVA,118,316;%%
\bibitem{glashow2}
V.~S.~Berezinsky and A.~Z.~Gazizov, {\it JETP\,Lett.} {\bf 25}, 254 (1977).
%%CITATION = ZFPRA,25,276;%%
\bibitem{glashow3}
K.~O.~Mikaelian and I.~M.~Zheleznykh, {\it Phys.\,Rev.} {\bf D22}, 2122 (1980).
%%CITATION = PHRVA,D22,2122;%%
\bibitem{dumand}
K.~K.~Young {\it et al.} (The {\sc Dumand} Collaboration), in Proceedings of
Joint 15th International Lepton Photon Symposium at High Energies and European
Physical Society Conference on High-energy Physics, Geneva, Switzerland, 
25 July - 1 August 1991, Vol.\,1, p.\,662. See also URL 
{\it http://www.phys.hawaii.edu/dmnd/dumand.html}
\bibitem{baikal_1}
L.~B.~Bezrukov {\it et al.} (The {\sc Baikal} Collaboration), e-print
{\it astro-ph/9601161}; V.~A.~Balkanov {\it et al.} (The {\sc Baikal}
Collaboration), {\it Astropart.\,Phys.} {\bf 12}, 75 (1999) (available from 
$<$arXiv:{\it astro-ph/9903341}$>$).
%%CITATION = ASTRO-PH 9601161;%%
%%CITATION = ASTRO-PH 9903341;%%
\bibitem{amandab}
P.~O.~Hulth {\it et al.} (The {\sc Amanda} Collaboration), e-print
{\it astro-ph/9612068}.
%%CITATION = ASTRO-PH 9612068;%%
\bibitem{amandaa}
P.~Askebjer {\it et al.} (The {\sc Amanda} Collaboration),
{\it Nucl.\,Phys.\,Proc.\,Suppl.} {\bf 38}, 287 (1995).
%%CITATION = NUPHZ,38,287;%%
\bibitem{backgr}
P.~Amram {\it et al.} (The {\sc Antares} Collaboration), {\it Astropart. Phys.}
{\bf 13}, 127 (2000) (available from $<$arXiv:{\it astro-ph/9910170}$>$).
%%CITATION = ASTRO-PH 9910170;%%
\bibitem{sediment}
P.~Amram {\it et al.} (The {\sc Antares} Collaboration), 
{\it Astropart.\,Phys.} {\bf 19}, 253 (2003) (available from 
$<$arXiv:{\it astro-ph/0206454}$>$).
%%CITATION = ASTRO-PH 0206454;%%
\bibitem{optic}
J.~A.~Aguilar  {\it et al.} (The {\sc Antares} Collaboration), e-print
{\it astro-ph/0412126} (to be published in {\it Astropart. Phys.}).
%%CITATION = ASTRO-PH 0412126;%%
\bibitem{prot}
M.~Circella {\it et al.} (The {\sc Antares} Collaboration), in Proceedings of
28th International Cosmic Ray Conference (ICRC 2003), Tsukuba, Japan, 
31 July - 7 August 2003, p.\,1529.
\bibitem{OM}
P.~Amram {\it et al.} (The {\sc Antares} Collaboration), {\it NIM} {\bf A484},
369 (2002) (available from $<$arXiv:{\it astro-ph/0112172}$>$).
%%CITATION = ASTRO-PH 0112172;%%
\bibitem{PMT}
J.~A.~Aguilar  {\it et al.} (The {\sc Antares} Collaboration), to be 
submitted to {\it NIM}.
\bibitem{data}
M.~C.~Bouwhuis {\it et al.} (The {\sc Antares} Collaboration), in Proceedings 
of 28th International Cosmic Ray Conference (ICRC 2003), Tsukuba, Japan, 
31 July - 7 August 2003, p.\,1541.
\bibitem{lms}
A.~L'Abbate, T.~Montaruli and I.~Sokalski, e-print {\it hep-ph/0406133} (to be 
published in {\it Astropart. Phys.}).
%%CITATION = HEP-PH 0406133;%%
\bibitem{cang}
R.~Enomoto {\it et al.} (The {\sc Cangaroo} Collaboration), {\it Nature},
{\bf 416}, 823 (2002).
%%CITATION = NATUA,416,823;%%
\bibitem{MACRO_pls}
M.~Ambrosio {\it et al.} (The {\sc Macro} Collaboration), {\it Astrophys.\,J.} 
{\bf 546}, 1038 (2001) (available from $<$arXiv:{\it astro-ph/0002492}$>$).
%%CITATION = ASTRO-PH 0002492;%%
\bibitem{AMANDA_pls}
M.~Ackermann {\it et al.} (The {\sc Amanda} Collaboration), e-print
{\it astro-ph/0412347} (submitted to {\it Phys.\,Rev.\,D}.)
%%CITATION = ASTRO-PH 0412347;%%
\bibitem{AMANDA_dif}
K.~Woschnagg {\it et al.} (The {\sc Amanda} Collaboration), to appear in
Proceedings of 21st International Conference on Neutrino Physics and
Astrophysics (Neutrino 2004), Paris, France, 14-19 June 2004 (available from 
$<$arXiv:{\it astro-ph/0409423}$>$).
%%CITATION = ASTRO-PH 0409423;%%
\bibitem{IceCube_pls}
J.~Ahrens {\it et al.} (The {\sc IceCube} Collaboration), 
{\it Astropart.\,Phys.} {\bf 20}, 507 (2004) (available from 
$<$arXiv:{\it astro-ph/0305196}$>$).
%%CITATION = ASTRO-PH 0305196;%%
\bibitem{ANTARES_pls}
A.~Heijboer {\it et al.} (The {\sc Antares} Collaboration), in Proceedings of
28th International Cosmic Ray Conference (ICRC 2003), Tsukuba, Japan, 
31 July - 7 August 2003, p.\,1321.
\bibitem{distefano}
C.~Distefano,  D.~Guetta, E.~Waxman and A.~Levinson, {\it Astrophys.\,J.} 
{\bf 575}, 378 (2002) (available from $<$arXiv:{\it astro-ph/0202200}$>$).
%%CITATION = ASTRO-PH 0202200;%%
\bibitem{Baikal_nu}
Zh.-A.~M.~Djilkibaev {\it et al.} (The {\sc Baikal} Collaboration), to appear 
in Proceedings of 21st International Conference on Neutrino Physics and
Astrophysics (Neutrino 2004), Paris, France, 14-19 June 2004.
\bibitem{MACRO_nu}
M.~Ambrosio {\it et al.} (The {\sc Macro} Collaboration), 
{\it Astropart.\,Phys.} {\bf 20}, 1 (2003) (available from 
$<$arXiv:{\it astro-ph/0203181}$>$).
%%CITATION = ASTRO-PH 0203181;%%
\bibitem{frejus}
W.~Rhode {\it et al.} (The {\sc Frejus} Collaboration), {\it Astropart.\,Phys.}
{\bf 4}, 217 (1996).
%%CITATION = APHYE,4,217;%%
\bibitem{AMANDA_nu}
J.~Ahrens {\it et al.} (The {\sc Amanda} Collaboration),
{\it Phys.\,Rev.\,Lett.} {\bf 92}, 071102 (2004) (available from 
$<$arXiv:{\it astro-ph/0309585}$>$).
%%CITATION = ASTRO-PH 0309585;%%
\bibitem{wb}
E.~Waxmann and J.~N.~Bahcall, {\it Phys.\,Rev.} {\bf D59}, 023002 (1999)
(available from $<$arXiv:{\it hep-ph/9807282}$>$).
%%CITATION = HEP-PH 9807282;%%
\bibitem{mpr}
K.~Mannheim,  R.~J.~Protheroe and J.~P.~Rachen, {\it Phys.\,Rev.} {\bf D63}, 
023003 (2001) (available from $<$arXiv:{\it astro-ph/9812398}$>$).
%%CITATION = ASTRO-PH 9812398;%%
\bibitem{MACRO_neu}
M.~Ambrosio {\it et al.} (The {\sc Macro} Collaboration), {\it Phys.\,Rev.}
{\bf D60}, 082002 (1999) (available from $<$arXiv:{\it hep-ex/9812020}$>$).
%%CITATION = HEP-EX 9812020;%%
\bibitem{Baksan_neu}
O.~V.~Suvorova, in Proceedings of 2nd International Conference on Physics 
Beyond the Standard Model: Accelerator, Nonaccelerator and Space Approaches 
(Beyond the Desert 1999), Ringberg Castle, Tegernsee, Germany, 6-12 June 1999,
p.\,853 (available from $<$arXiv:{\it hep-ph/9911415}$>$).
%%CITATION = HEP-PH 9911415;%%
\bibitem{SK_neu}
S.~Desai {\it et al.} (The {\sc Super-Kamiokande} Collaboration),
{\it Phys.\,Rev.} {\bf D70}, 083523 (2004), {\bf erratum}: {\it ibid.} 
{\bf D70}, 109901 (2004) (available from $<$arXiv:{\it hep-ex/0404025}$>$).
%%CITATION = HEP-EX 0404025;%%
\bibitem{sugra}
A.~H.~Chamseddine, R.~Arnowitt and P.~Nath, {\it Phys.\,Rev.\,Lett.} {\bf 49},
970 (1982); R.~Barbieri, S.~Ferrara and C.~A.~Savoy, {\it Phys.\,Lett.} 
{\bf B119}, 343 (1982); L.~J.~Hall, J.~Lykken and S.~Weinberg, 
{\it Phys.\,Rev.} {\bf D27}, 2359 (1983). 
%%CITATION = PRLTA,49,970;%%
%%CITATION = PHLTA,B119,343;%%
%%CITATION = PHRVA,D27,2359;%%
\bibitem{km3}
See URL {\it http://www.km3net.org}
\end{thebibliography}
\end{document}